\documentclass[showpacs,preprintnumbers, twocolumn,
amsmath,amssymb,APSl,prd,nofootinbib,superscriptaddress]{revtex4-2}

\usepackage{amsmath}
\usepackage{amssymb}
\usepackage{amsfonts}
\usepackage{graphicx,bm}
\usepackage[colorlinks=true]{hyperref}
\usepackage{epsf}
\usepackage{enumerate}
\usepackage{hhline}
\usepackage{array}
\usepackage{tabularx}
\usepackage{color}
\usepackage[table,xcdraw]{xcolor}
\usepackage{subfigure}

\newcommand{\be}{\begin{equation}}
\newcommand{\ee}{\end{equation}}
\newcommand{\bea}{\begin{eqnarray}}
\newcommand{\eea}{\end{eqnarray}}
\newcommand{\beaa}{\begin{eqnarray*}}
\newcommand{\eeaa}{\end{eqnarray*}}

\newcommand{\nn}{\nonumber \\}

\def\be{\begin{equation}}
\def\ee{\end{equation}}
\def\bea{\begin{eqnarray}}
\def\eea{\end{eqnarray}}

\begin{document}
\title{General spherically symmetric black bounces within non-linear electrodynamics}

\author{G. Alencar}
\email{geova@fisica.ufc.br}
\affiliation{Department of Physics, Universidade Federal do Cear\'a (UFC), Campus do Pici, Fortaleza - CE, C.P. 6030, 60455-760 - Brazil}

\author{Albert Duran-Cabac\'es} \email{albert.duran22@uva.es}
\affiliation{Department of Theoretical Physics, Atomic and Optics, and Laboratory for Disruptive Interdisciplinary Science (LaDIS), Campus Miguel Delibes, \\ University of Valladolid UVA, Paseo Bel\'en, 7,
47011 - Valladolid, Spain}

\author{Diego Rubiera-Garcia} \email{drubiera@ucm.es}
\affiliation{Departamento de F\'isica Te\'orica and IPARCOS,
	Universidad Complutense de Madrid, E-28040 Madrid, Spain}

\author{Diego S\'aez-Chill\'on G\'omez}
\email{diego.saez@uva.es} 
\affiliation{Department of Theoretical Physics, Atomic and Optics, and Laboratory for Disruptive Interdisciplinary Science (LaDIS), Campus Miguel Delibes, \\ University of Valladolid UVA, Paseo Bel\'en, 7,
47011 - Valladolid, Spain}
\affiliation{Department of Physics, Universidade Federal do Cear\'a (UFC), Campus do Pici, Fortaleza - CE, C.P. 6030, 60455-760 - Brazil}

\begin{abstract}

Over the last years, the search of new regular black bounce solutions has drawn a lot of attentional over the international community working in gravitation. Indeed, in the era of gravitational waves detections out of binary mergers and of the imaging of the plasma around supermassive black holes, the study of everywhere regular solutions has become a common trend given the unique opportunity posed by multi-messenger astronomy to test deviations from the Kerr family of solutions. Among them, in this paper we consider the black bounce paradigm introduced by Simpson and Visser in [JCAP \textbf{02}, 042 (2019)], and provide a general procedure for reconstructing static spherically symmetric black bounce-type solutions that might interpolate between regular black holes and wormholes. We show that even after imposing some smoothness and flatness conditions on the metric components, additional analysis is required to obtain a well-defined black bounce solution. Then, the corresponding matter Lagrangian is reconstructed by using non-linear electrodynamics and the energy conditions are studied.

\end{abstract}
%
%
%\pacs{04.50.Kd, 98.80.-k, 95.36.+x}
%%
%%
\maketitle
%
%%%%%%%%%%%%%%%%%%%%%%%%%%%%%%%%%%%%%
\section{Introduction}
%%%%%%%%%%%%%%%%%%%%%%%%%%%%%%%%%%%%%

Different regular compact objects solutions have drawn a lot of attention in the literature and new solutions have been proposed over the last years. The detection of gravitational wave signals from coalescing binary systems (either two black holes \cite{LIGOScientific:2016aoc} or a black hole and a neutron star \cite{LIGOScientific:2017vwq}), together with the imaging of the shape of the surrounding plasma around the central supermassive objects (likely supermassive black holes) at the heart of M87 \cite{EventHorizonTelescope:2019dse} and our own Milky Way galaxies \cite{EventHorizonTelescope:2022wkp}, has marked the beginning of the {\it multimessenger era}, namely, astronomy with different carriers (light, gravitational waves, neutrinos, and cosmic rays). Within this era we have an unique opportunity to test the existence and properties of any compact object, eventually allowing us to peer into the behaviour of the gravitational interaction on its strong-field regime \cite{Addazi:2021xuf}.

Regarding black holes, General Relativity (GR) provides us with a single solution in vacuum, as given by the well-known family of Kerr solutions, completely described by two parameters: the mass $M$, and the angular momentum $J$  (see for instance, Ref.~\cite{Hobson:2006se}), while a charge $Q$ can be added to extend the family to the Kerr-Newman one though such a charge is typically regarded as negligible in astrophysical scenarios. This family arises under very general and natural conditions within the full gravitational collapse of a fuel-exhausted, massive-enough star. However, such a process also holds the demise of the very theory (GR) it nurtures black holes: the development of a space-time singularity on its interior regions, as given by the incompleteness of some geodesics paths there, is unavoidable \cite{Senovilla:2014gza}. Given the fact that no other solution (in vacuum) is possible within GR due to the uniqueness theorems, one is faced with the dilemma of having singular solutions which nonetheless are perfectly consistent with every astrophysical observation made so far, or accept that there should be regular solutions beyond what theorems account for. While it has been generally argued that the resolution of space-time singularities should lie on a complete description of GR and/or gravitation at high-energy regimes, where the quantum effects of gravity should manifest to save the day, this does not prevent us from exploring classical ways to construct regular solutions that might actually reflect the effects of such slippery theory of quantum gravity at lower energy scales.

Hence, the quest for regular black hole solutions that provide similar dynamics for light and GW trajectories and consistence with observations as the Kerr black holes of GR, but being absent of any space-time singularity, has become a trend in the literature of the field (for a review on regular solutions, see Ref.~\cite{Bambi:2023try}). Due to the uniqueness (in vacuum) of the Kerr black hole within GR, any such regular solution must either i) include additional matter fields (see e.g. \cite{Herdeiro:2015waa}), ii) be found within theories beyond GR, iii) or both of them. Prominent works within the first path include the regular black holes by Bardeen \cite{bardeen:1968non}, Hayward \cite{Hayward:2005gi} and Ay\'on-Beato and Garcia \cite{Ayon-Beato:1998hmi}, sourced by non-linear electrodynamics (NEDs), or the Morris-Thorne space-time \cite{Morris:1988cz} pioneering the wormhole physics field \cite{VisserBook}. For the sake of this work we bring forward the Simpson-Visser (SV) proposal \cite{Simpson:2018tsi} or, as it is more commonly known, the {\it black bounces} (BB). 

The BB proposal combines a Schwarzschild space-time with a bounce in the radial-like coordinate of the Morris-Thorne form to prevent the focusing of geodesics and make the core of the object regular. The object built this way may either contain horizons, in whose case they represent regular black holes, or not, yielding a horizonless regular object interpreted as a traversable wormhole. This proposal was later extended towards other bouncing behaviours  \cite{Lobo:2020ffi,Tsukamoto:2021caq}, charged BBs \cite{Franzin:2021vnj}, axially symmetric \cite{Islam:2021ful,Ghosh:2022mka} and cylindrically symmetric (black strings) \cite{Lima:2022pvc,Bronnikov:2023aya,Lima:2023arg,Lima:2023jtl} generalizations. Geodesic lensing and their observables have been widely studied in order to reveal differences of these solutions as compared to canonical black hole ones \cite{Nascimento:2020ime,Tsukamoto:2020bjm,Zhang:2022nnj,Guo:2021wid}.

All of these implementations of the BB idea have in common that require matter sources that, within the GR framework, violate the null energy condition, in agreement with the hypothesis underlying the singularity theorems. In addition, the radial and tangential components of the pressure holding such sources turn out to be different according to the Einstein tensor, such that different sources are required in order to reconstruct this type of solutions. In this sense, non-linear electrodynamics (NEDs) together with a scalar field provide a way for reconstructing BBs space-times \cite{Rodrigues:2023vtm, Bronnikov:2021uta,Alencar:2024yvh}. Other recent papers explore the possibility of reconstructing and analyzing the structure of spherically symmetric black bounces with anisotropic fluids \cite{Lessa:2024erf}.
 
 The purpose of this paper is to reconstruct general BB solutions with spherical symmetry that describe regular space-times. To do so, we consider general spherically symmetric space-times where the appropriate conditions are imposed on the metric functions to get bound in the radial function and a regular BB solution. In addition, several constraints are analyzed and the corresponding curvature invariants are obtained in order to get a reliable reconstruction of a regular space-time. Then, the SV space-time is generalized by promoting the original metric with two independent parameters to one with four independent parameters. The corresponding energy-momentum tensor is obtained for the general case as for the generalization of  SV space-time. Different properties of the dynamics of the space-times are analysed, including null geodesics and the energy conditions.
 % which are violated. 
 
 The paper is organized as follows: section \ref{general} is devoted to introduce a general spherically symmetric metric that satisfies some particular conditions and different cases are analyzed. Then, in section \ref{generalSV}, such a procedure is applied to a generalisation of the Simpson-Visser metric. In section \ref{reconstruction}, the corresponding matter Lagrangian is obtained through non-linear electrodynamics and a scalar field, while energy conditions are analyzed in section \ref{energy}. Finally, section \label{conclusions} gathers the conclusions of the paper.

%%%%%%%%%%%%%%%%%%%%%%%%%%%%%%%%%%%%%
\section{General spherically symmetric black bounce spacetimes}
\label{general}
%%%%%%%%%%%%%%%%%%%%%%%%%%%%%%%%%%%%%
Over the years, two kinds of solutions have generally described spherically symmetric space-times. The first one is the typical black hole solution, given by 
\begin{equation}\label{BH}
 ds^2=-A(r) dt^2 +A^{-1}(r) dr^2 + r^2 d\Omega^2  \ , 
\end{equation}
where $d\Omega^2 = d \theta^2 + \sin^2 \theta d  \varphi^2$ is the line element in the two-spheres. From the above, we can obtain the Schwarzschild solution, in vacuum, with a singularity at $r=0$ and horizons at the roots  $A(r=r^a)=0$. For regular black holes, the presence of matter sources may avoid the singularity. 

%In the last case, regularity conditions must be satisfied.  

The second solution is the Morris-Thorne (MT) wormhole solution, given by
\be\label{WR}
ds^2=-A(r) dt^2 + dr^2 + C^2(r) d\Omega^2. 
\ee
For $A=1$ and $C=\sqrt{r^2+a^2}$, we get the simplest Ellis-Bronikov regular solution with the throat at $r=0$. For the general case, Morris and Thorne proposed some conditions, which we follow here. First, we assume that $-\infty<r<\infty$ to generate a bounce in the radial function $C^2(r)$ whose throat can be located at $r=0$, without any loss of generality.  Since $\sqrt{-g}=|C|\sqrt{A}$,  we must impose $A(r),C(r)\neq 0$ and $A(r),C(r)\neq \infty$ in the whole domain to have a well-defined volume element and to display a smooth behavior at $r=0$.  At such a point, the radius of the two spheres turns out finite and non-null, implementing the bounce via the conditions
\be
C(r=0)=c \quad ; \quad C'(r=0)=0 \quad ; \quad C''(r=0)<0
\label{throatR}
\ee
realizing the minimum character of this point. This way, we construct a manifold $\mathcal{M} = \mathcal{M}_+ \cap \mathcal{M}_-$ where $\mathcal{M}_{\pm}$ are the two space-times on each side of the bounce. Finally, we can impose that both metrics are asymptotically flat. For the metric (\ref{BH}) we just need that $A(\pm \infty)=1$ and for the metric (\ref{WR}) we impose
\be
\text{For}\quad r\rightarrow\pm\infty\ ,  \quad  A\rightarrow 1 \quad  C\rightarrow r\ .
\label{FlatCond}
\ee

In this direction, in 2019, Simpson and Visser proposed a black-bounce by implementing the transformation
\begin{equation}
r\to\sqrt{r^2+a^2}
\end{equation} 
in the black hole solution  (\ref{BH}) to obtain \cite{Simpson:2018tsi}
\be\label{SV}
 ds^2=-A(\sqrt{r^2+a^2}) dt^2 +\frac{1}{A(\sqrt{r^2+a^2})}dr^2 + (r^2+a^2) d\Omega^2  \ .
\ee
With this prescription, the SV solution interpolates between a black hole and a wormhole, depending on the value of the parameter $a$. Some general conclusions can be obtained directly. First, the volume integral is always well defined since $\sqrt{-g}=r^2+a^2$. Second, the singularity is avoided since the area of the two spheres $S=4\pi C^2(r)$ is always non-vanishing.. Finally, the horizons are now located at $r_h=r_a^2-a^2$, where $r_a$ are the roots of the original black hole solution. Therefore, for $a<r_a$, we have a regular black hole, and for $a>r_a$, we have a wormhole. Therefore, the horizons are avoided if the wormhole throat is located beyond  $r_h$, namely, if $a>r_h$.

However, a more general black bounce can be achieved by considering a spherically symmetric space-time metric:of the form
\be
ds^2=-A(r) dt^2 +B^{-1}(r) dr^2 + C^2(r) d\Omega^2  \ ,
\label{GeneralBB}
\ee 
where the bounce is implemented through (\ref{throatR}). On the other hand, there might be roots where $A(r=r^{a})=0$ and/or $B(r=r^{b})=0$, which may lead to surfaces of infinite redshift and/or event horizons. The resulting space-time will be properly interpreted on a case-by-case basis, a main goal of our work. 

On the other hand, to further interpret the corresponding geometries encoded in the space-time metric (\ref{GeneralBB}), we shall analyze its geodesic structure. For radial geodesics, the metric turns out:
\be
ds^2=-A(r) dt^2 +B^{-1}(r) dr^2\ ,
\label{radialNull1}
\ee
where $ds^2=0$ is required for null trajectories. The complete set of the geodesic equations can be obtained through the Lagrangian density of a point particle, written as
\be
L=g_{\mu\nu}\dot{x}^{\mu}\dot{x}^{\nu}=-A(r)\dot{t}^2+B^{-1}(r) \dot{r}^2\ ,
\label{LagrangianGeo}
\ee
where an overdot refers to derivatives with respect to an affine parameter $\sigma$ over the particle's trajectory. Then, the corresponding geodesic equations lead to:
\bea
\frac{d}{d\sigma}\left(A(r)\dot{t}\right)&=&0\ , \label{geodesicsEqs1} \\
\ddot{r}+\frac{A'B}{2}\dot{t}^2-\frac{B'}{2B}\dot{r}^2&=&0\ .
\label{geodesicsEqs2}
\eea
The first equation can be directly integrated, which yields $A(r)\dot{t}=E$, where $E$ is a constant that follows from the time-like Killing vector associated to static metrics and which can be identified with the conservation of the energy. In addition, from (\ref{LagrangianGeo}) one can easily get a first integral. For massive particles, this first integral leads to:
\be
\dot{r}^{2}=\frac{B}{A} E^2-B\ .
\label{massiveRadial}
\ee
whereas for null radial geodesics, it reads instead
\be
\dot{r}^{2}=\frac{B}{A} E^2\ .
\label{radialNull2}
\ee
Moreover, it is also useful to analyze the behaviour of the tangent vector in the $(t,r)$ diagram for null geodesics, which is given by:
\be
\left(\frac{dt}{dr}\right)^2=\frac{1}{A(r) B(r)}\ ,
\label{tangent1}
\ee
Then, this tangent vector diverges at $A=0$ and/or $B=0$, which might reveal the presence of event horizons. Moreover, non-radial null geodesics might also provide important information about the space-time structure. In such a case, there is an extra equation to the system (\ref{geodesicsEqs1}) and (\ref{geodesicsEqs2}) corresponding to:
\bea
\frac{d}{d\sigma}\left(C^2(r)\dot{\varphi}\right)=0\ ,
\label{NonRadialgeodesicsEqs1}
\eea
where we have assumed $\theta=\pi/2$ with no loss of generality due to the spherical symmetry of the space-time. Then, Eq.(\ref{NonRadialgeodesicsEqs1}) can be directly integrated, leading to $C^2(r)\dot{\varphi}=L$, which just shows the conservation of the angular momentum. Hence, by combining the equations (\ref{geodesicsEqs1}) and (\ref{NonRadialgeodesicsEqs2}), the following equation is obtained for the radial coordinate:
\be
\dot{r}^2=\frac{B(r)}{A(r)}\left(\frac{E^2}{L^2}-\frac{A(r)}{C^2(r)}\right)\ ,
\label{NonRadialgeodesicsEqs2}
\ee
where we have absorbed a factor $L^2$ in the definition of the affine parameter. 

It should be stressed, however, that the three  metric components appearing in the line element (\ref{GeneralBB})  can actually be reduced to two independent ones by a suitable change of coordinates. To implement it, let us consider the following transformation of the radial coordinate:
\be
dx^2=\frac{A(r)}{B(r)}dr^2\ .
\label{coordChang}
\ee
so that the line element (\ref{GeneralBB}) becomes
\be
ds^2=-A(r(x)) dt^2 +A^{-1}(r(x)) dx^2 + C^2(r(x)) d\Omega^2 \ ,
\label{GeneralBB1}
\ee
where $x$ will be defined in a particular domain which might differ from the one of $r$, as provided by the integral (\ref{coordChang}). In the above coordinate, we also see that $\sqrt{-g}=C(x)$ and the volume integral is always well-defined. 

Finally, by assuming that Einstein field equations hold, then $G^{t}_{t}=G^{r}_{r}$ must be imposed as one crosses the horizon (if any) in order to ensure that $\rho$ is continuous. Therefore, we must  have 
\begin{equation}
    2\frac{C''}{C}\Big \vert_{r_h}=\frac{C'^{2}}{C^{2}}\Big \vert_{r_h}.
\end{equation}
Nevertheless, for our purposes, we shall analyze the different cases depending on the roots of the functions $A$ and $B$ and make use of the transformation (\ref{GeneralBB}) when necessary to clarify the physical interpretation of each scenario.

%%%%%%%%%%%%%%%%%%%%%%%%%
\subsection{Case I: $A(r)\neq0$ and $B(r)\neq0$.}
%%%%%%%%%%%%%%%%%%%%%%%%%

In this case, the space-time metric (\ref{GeneralBB}) together with the conditions (\ref{FlatCond}) and (\ref{throatR}) describes a two-way wormhole with  presence neither of event horizons nor of singularities. The latter comes from the assumed bouncing behaviour of the function $C^2(r)$, which prevents the focusing of geodesics in Eq.(\ref{NonRadialgeodesicsEqs2}). This is the case of most wormhole solutions proposed in the literature. 

%%%%%%%%%%%%%%%%%%%%%%%%%
\subsection{Case II: $A(r)\neq0$ and $B(r=r_{\pm}^{b})=0$.}
%%%%%%%%%%%%%%%%%%%%%%%%%

Let us assume now that $B(r=r_{\pm}^{b})=0$ has two real roots, one of each side of the bounce, i.e., on $\mathcal{M}_{\pm}$ (we do not consider more than one such root since we are not interested here on inner horizons). It is clear that the space-time metric (\ref{GeneralBB}) is not well defined at $r=r_{\pm}^{b}$ but the change of coordinates (\ref{coordChang}) removes this problem such that it is just a coordinate singularity. Does the surfaces $r_{\pm}$ represent event horizons? (on each side of the bounce). A glance at the radial null equation (\ref{radialNull2}) reveals  that one photon might reach the hypersurfaces $r=r_{\pm}^{b}$ but is not allowed to keep going below, since the tangent vector  (\ref{radialNull2}) does not exist within the region $r_{-}^{b}<r<r_{+}^{b}$. However, by the transformation of coordinates (\ref{coordChang}), the metric turns out completely regular (\ref{GeneralBB}), so it seems that such non-accessible region is just an artefact of a bad choice of coordinates. The only question is whether $B(r)$ (and $A(r)$) are well defined within a continuous domain for $x$ over all the space. This is actually the case as far as $r=r_{\pm}^{b}$ are roots of first order in $B(r)$. For $r>r_{+}^{b}$, the integration of the transformation of coordinates (\ref{coordChang}) yields:
\be
x=\int^{r}_{r_{+}^{b}} dr'\sqrt{\frac{A}{B}}\ ,
\ee
which might be expressed in a differential form as follows:
\be \label{xI}
x'(r)-\sqrt{\frac{A}{B}}=0\ , \quad x(r_{+}^{b})=0 \ .
\ee
One can proceed in the same way for $r<r_{+}^{b}$. Hence, both roots are matched at $x=0$ such that the non-accesible region $r_{-}^{b}<r<r_{+}^{b}$ is just the result of a bad choice of the radial coordinate $r$ whose domain does not cover such range of values. Moreover, by the flatness condition (\ref{FlatCond}), one has:
\be
\text{For}\quad r\rightarrow\infty\ \quad \rightarrow\quad x\sim r\ .
\ee
Finally, by using (\ref{xI}) we note that 
\begin{equation}
   \frac{dC(x)}{dx}\Big \vert_{x=0}=\left[C'(r)\sqrt{\frac{B}{A}}\right]_{r=r^b}=0.
\end{equation}
Therefore, the metric as expressed in terms of $x$ is regular everywhere, $-\infty<x<\infty$, and no event horizons arise. Hence, one might conclude that the metric (\ref{GeneralBB}) represents a two-way traversable wormhole with the throat located at $x=0$ whose areal radius is given by $4\pi C^2(x=0)$. 

%%%%%%%%%%%%%%%%%%%%%%%%%
\subsection{Case III: $A(r=r_{\pm}^{a})=0$ and $B(r)\neq0$.}
%%%%%%%%%%%%%%%%%%%%%%%%%

For this case, where $A(r=r_{\pm}^{a})=0$ has two real roots, also symmetric with respect to $r=0$, there seems to be apparently again a non-accesible region for $r_{-}^{a}<r<r_{+}^{a}$, where the geodesics equations (\ref{radialNull2}) and (\ref{massiveRadial}) are not defined. Furthermore, the geodesic equations contain a divergence at $r=r_{\pm}^{a}$. In addition, the space-time metric expressed in terms of the radial coordinate $x$ keeps singular at $x_{\pm}^{a}=x_{\pm}^{a}(r_{\pm}^{a})$, as shown by (\ref{GeneralBB}). However, this singularity might be removed by performing an additional change of coordinates. Radial null geodesics for the metric (\ref{GeneralBB}) are given by:
\be
ds^2=-A(r(x)) dt^2 +A^{-1}(r(x)) dx^2=0
\ee
so that we can introduce the coordinate change
\be 
dt^2=A^{-1}(r(x)) dx^2=dx^{*2}\ ,
\ee
where we have introduced a new coordinate $x^{*}$ analog to the usual Regge-Wheeler coordinate for the Schwarzschild space- time. Then, radial null geodesics are given by $d(t\mp x^{*})=0$ and we can introduce another coordinate, the so-called ingoing radial null coordinate:
\be
v=t+x^{*}\ .
\ee
Finally, the metric (\ref{GeneralBB}) becomes:
\be
ds^2=-A(x)dv^2+dvdx+C^2(x)d\Omega^2\ ,
\ee
which is clearly regular at $x=x_{\pm}^{a}$, where such hypersurfaces are null. Do they form two separate event horizons? If we follow a similar procedure as in the previous case, the change of coordinates for $r>r_{+}^{a}$ is given by:
\be
x=\int^{r}_{r_{+}^{a}} dr'\sqrt{\frac{A}{B}}\ ,
\ee
i.e., formally similar as in the previous case, and which is again expressed in differential form as
\be
x'(r)-\sqrt{\frac{A}{B}}=0\ , \quad x(r_{+}^{a})=0 \ .
\ee
One can proceed in the same way for $r>r_{-}^{a}$ and find that $x(r_{-}^{a})=0$ and,  hence, there might be a unique event horizon at $x=0$. However, as done before, we have
\begin{equation}
   \frac{dC(x)}{dx}\Big \vert_{x=0}=\left[C'(r)\sqrt{\frac{B}{A}}\right]_{r=r^a}=\infty.
\end{equation}
and the point $x=0$ thus cannot be a throat. We conclude this case cannot describe a black bounce. 

%%%%%%%%%%%%%%%%%%%%%%%%%
\subsection{Case IV: $A(r=r_{\pm}^{a})=0$ and $B(r=r_{\pm}^{b})=0$.}
%%%%%%%%%%%%%%%%%%%%%%%%%

This case reduces just to one of the above cases. It just depends whether $|r_{\pm}^{a}|<|r_{\pm}^{b}|$ (case II) or $|r_{\pm}^{a}|\geq|r_{\pm}^{b}|$ (case III). However, as shown above, the only allowed possibility is  $|r_{\pm}^{a}|<|r_{\pm}^{b}|$.

%%%%%%%%%%%%%%%%%%%%%%%%%
\subsection{Case V: $A(r=r_{\pm}^{a})=0$ and $B(r=r_{\pm}^{a})=0$ }
%%%%%%%%%%%%%%%%%%%%%%%%%
Here, the metric (\ref{GeneralBB}) might be well defined for $-\infty<r<\infty$, depending on the relative order of the roots of $A(r)$ and $B(r)$. There are two event horizons located at $r=r_{\pm}^{a}$, i.e., on each side of the bounce. This case represents a regular black hole. Together with the results of case IV, the only possibility is that 
\begin{equation}
    r_b^{+}\geq r_a^+.
\end{equation}

%%%%%%%%%%%%%%%%%%%%%%%%%%%%%%%%%%%%%
\section{A particular case: generalization of Simpson-Visser space-time}
\label{generalSV}
%%%%%%%%%%%%%%%%%%%%%%%%%%%%%%%%%%%%%

Let us apply the above analysis for a particular space-time metric that represents a kind of the generalization of SV solution. To do it so, the original metric of \cite{Simpson:2018tsi} is promoted to contain three independent free parameters as given the line element (\ref{GeneralBB}) with the functions
\bea
A(r)&=&1-\frac{2m}{\sqrt{r^2+a^2}}, \quad B(r)=1-\frac{2m}{\sqrt{r^2+b^2}}, \label{eq:SV1}\\
C^2(r)&=&r^2+c^2  \label{eq:SV2}
\eea
It is straightforward to note that this metric coincides with the usual SV metric except for the fact that each diagonal element has its own independent free parameter (besides $m$), labelled $a$, $b$ and $c$. Therefore, we can recover the SV metric just by setting $a=b=c$, while the choice $a=b=c=0$ recovers Schwarzchild space-time. In addition, the metric accomplishes the requirement (\ref{FlatCond}) to be asymptotically flat  and the throat of the possible wormhole located at $r=0$ has a topology $S^2$ with an areal radius $4\pi R^2=4 \pi c^2$.

As pointed out in the previous section, the tangent vector (\ref{tangent1}) diverges for $A(r)=0$ and/or $B(r)=0$, whose roots can be easily obtained in this case as
\bea
A(r)=0\ \quad \rightarrow \quad r_{\pm}^{a}=\sqrt{4m^2-a^2}\ , \nn
B(r)=0\ \quad \rightarrow \quad r_{\pm}^{b}=\sqrt{4m^2-b^2}\ .
\label{rootsAB}
\eea
Let us now analyze each case as above.

%%%%%%%%%%%%%%%%%%%%%%%%%
\subsection{Case I: $a>2m\ , b>2m$}
%%%%%%%%%%%%%%%%%%%%%%%%%

In this case there are no real roots in (\ref{rootsAB}), such that there are no possible event horizons and the metric describes a two-way traversable wormhole with a throat of areal radius $S=4\pi c^2$ with a good behaviour in the radial coordinate for all its domain $-\infty<r<\infty$. One might change to the coordinate $x$ as given in (\ref{coordChang}) and re-express the metric as in Eq.(\ref{GeneralBB}), where the metric reduces to three independent free parameters ($m$, $a$, and $b$), resembling in its behaviour the original SV space-time for the case of a two-way traversable wormhole.

%%%%%%%%%%%%%%%%%%%%%%%%%
\subsection{Case II: $a>2m\ , b\leq2m$}
\begin{figure}[t!]
   \centerline{ \includegraphics[scale=0.7,trim=5mm 0mm -2mm 0mm]{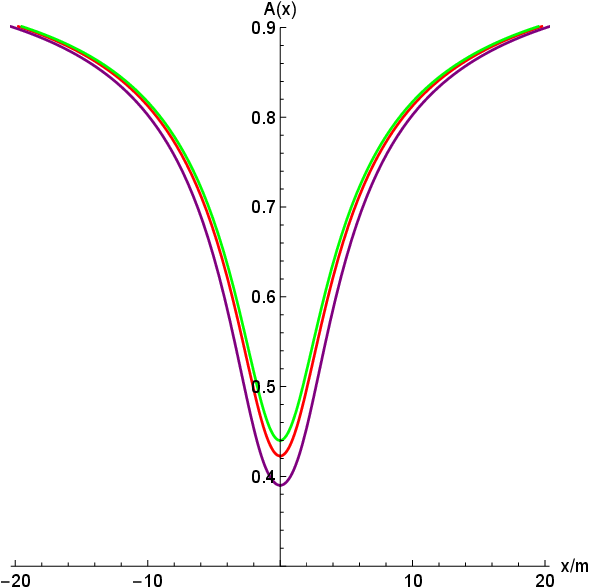}}
\caption{The metric component $A(x)$ for $a/m=3$ and $b/m=3/2$ (purple), $b/m=1$ (red) and $b/m=1/2$ (green). As shown, the metric is well-behaved in all the domain of $x$ and no horizons arise. This case represents a two-way traversable wormhole.}
  \label{F1}
\end{figure}

%%%%%%%%%%%%%%%%%%%%%%%%%
In this case we have solutions for which $B(r_{\pm}^{b})=0$, which would seem to represent null hypersurfaces located at $r=r_{\pm}^{b}$ with the tangent vector (\ref{tangent1}) diverging there. However, by performing the change of coordinates (\ref{coordChang}), one leads to the equation:
\be
x'(r)-\sqrt{\frac{1-\frac{2m}{\sqrt{r^2+a^2}}}{1-\frac{2m}{\sqrt{r^2+b^2}}}}=0\ , \quad x(r_{+}^{b})=0 \ ,
\ee
This way the metric turns out into the form (\ref{GeneralBB}) and it is regular at $x=0$. The new radial coordinate is well-defined for $-\infty<x<+\infty$. Moreover, the metric component $A(x)$ behaves smoothly everywhere, as shown in Fig.~\ref{F1} for a particular choice of the free parameters and the spacetime corresponds to a two-way traversable wormhole. Hence, the  apparent troubles that we had from the original metric coefficients where just due to a bad choice of the radial coordinate $r$. 

\begin{figure}[t!]
   \centerline{ \includegraphics[scale=0.7,trim=5mm 0mm -2mm 0mm]{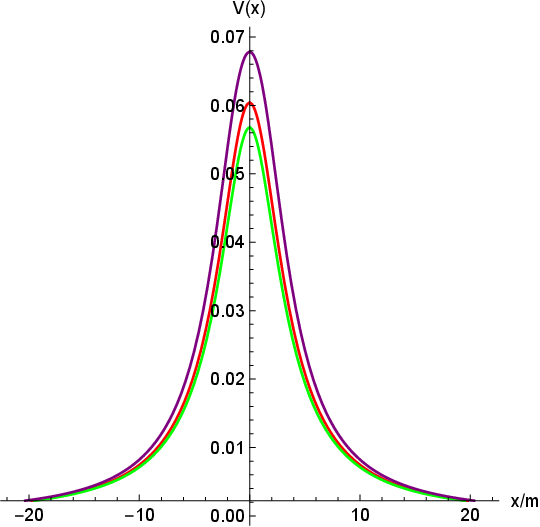}}
\caption{The effective potential $V(x)$ for $a/m=3$ and $b/m=3/2$ (purple), $b/m=1$ (red) and $b/m=1/2$ (green). As shown, an unstable circular orbit for photons (i.e. a maximum in the effective potential) arises at $x=0$.}
  \label{F1a}
\end{figure}

On the other hand, one can infer the behaviour of null geodesics by analyzing Eq.(\ref{NonRadialgeodesicsEqs2}), which in terms on the radial coordinate leads to:
\be
\dot{x}^2=\frac{E^2}{L^2}-V(x)\ ,
\label{PotentialEq}
\ee
where 
\begin{equation}
V(x)=\frac{A(x)}{C^2(x)}
\end{equation}
is the effective potential. The shape of this potential is depicted in Fig.~\ref{F1a}, where the presence of an unstable circular orbit (as given by a maximum in such a potential) at $x=0$ is evident. However, the effective potential is well-defined everywhere and so are the corresponding null trajectories in the whole range $x \in (-\infty,+\infty)$. 

%%%%%%%%%%%%%%%%%%%%%%%%%
\subsection{Case III: $a\leq2m$\ , $b>2m$}
\begin{figure}[t!]
   \centerline{ \includegraphics[scale=0.7,trim=5mm 0mm -2mm 0mm]{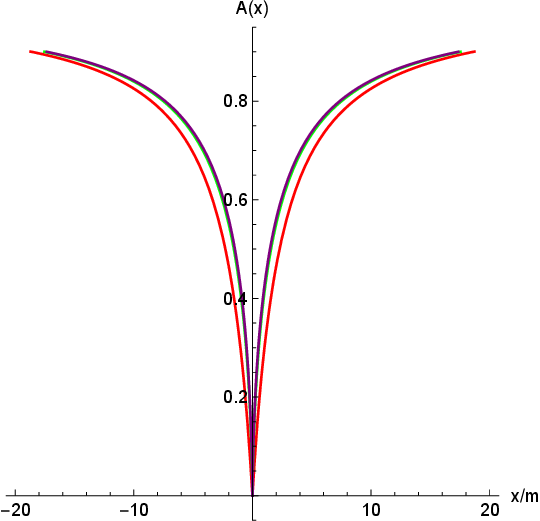}}
\caption{The metric component $A(x)$ for $a/m=0.5$ (purple), $a/m=1$ (green) and $a/m=2$ (red) and $b/m=3$. As shown, the metric exhibits a conic singularity at $x=0$. }
  \label{F2}
\end{figure}

\begin{figure}[t!]
   \centerline{ \includegraphics[scale=0.7,trim=5mm 0mm -2mm 0mm]{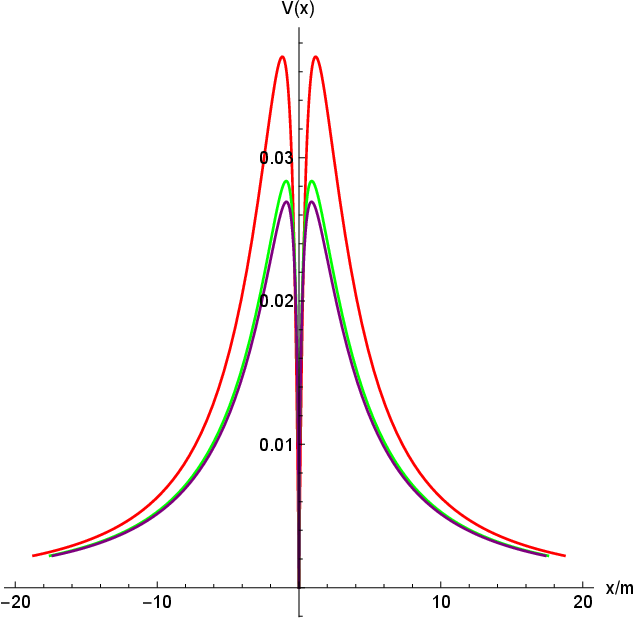}}
\caption{The effective potential $V(x)$ for $a/m=0.5$ (purple), $a/m=1$ (green) and $a/m=2$ (red) and $b/m=3$. As shown, an unstable circular orbit for photons arises at both sides of the throat.}
  \label{Fig2a}
\end{figure}

%%%%%%%%%%%%%%%%%%%%%%%%%
In this case $A(r_{\pm}^{a})=0$ and the tangent vector (\ref{tangent1}) diverges. However, by applying again the change of coordinates (\ref{coordChang}), one arrives to the metric (\ref{GeneralBB}) which is still singular at $x=0$ but can be regularized by following the procedure depicted in the previous section. Nevertheless, $A'(x)$ is not continuous at $x=0$, such that a conic singularity arises and geodesics (\ref{geodesicsEqs1}) are not well defined at that point, where $C'(x)$ diverges. The component of the metric $A=A(r(x))$ is depicted in Fig.~\ref{F2}. Moreover, the effective potential from the geodesic equation (\ref{PotentialEq}) is shown in Fig~\ref{Fig2a}, where two unstable circular orbits for photons arise but the derivative of the potential is not well defined at $x=0$. Hence, this case provides a metric with a conic singularity at $x=0$.

%%%%%%%%%%%%%%%%%%%%%%%%%
\subsection{Case IV: $a= b$}
%%%%%%%%%%%%%%%%%%%%%%%%%

Here the metric reduces to the usual SV one but the throat has an area radius $4 \pi c^2$. This case represents a traversable, two-way wormhole as far as $a=b>2m$, a regular black hole for $a=b<2m$ with two horizons located at $r=\pm\sqrt{4m^2-a^2}$ (i.e. on each side of the throat), or a one-way wormhole with an extremal null throat for $a=b=2m$.

%%%%%%%%%%%%%%%%%%%%%%%%%%%%%%%%%%%%%
\section{Reconstructing general black bounces in non-linear electrodynamics}
\label{reconstruction}
%%%%%%%%%%%%%%%%%%%%%%%%%%%%%%%%%%%%%

\subsection{General approach}

Our aim here is to obtain a stress-energy tensor that generates the modified SV space-time. With this purpose, we will follow the procedure in \cite{Alencar:2024yvh} very closely. We start considering the Hilbert-Einstein action in the presence of a scalar field and a NED Lagrangian, written as
\begin{equation}
S=\int \sqrt{-g} d^4 x\left[\frac{R}{2 \kappa^2}+\epsilon g^{\mu \nu} \partial_\mu \phi \partial_\nu \phi-V(\phi)-L(F)\right]
\label{NEDaction}
\end{equation}
where $g$ is the determinant of the space-time metric $g_{\mu \nu};$ $\phi$ is the scalar field with $V(\phi)$ its potential; $L(F)$ is the NED Lagrangian built as an arbitrary function of the invariant $F=\frac{1}{4} F^{\mu \nu} F_{\mu \nu}$, where $F_{\mu \nu}=\partial_\mu A_\nu-\partial_\nu A_\mu$ is the electromagnetic field tensor. Note that we have included a constant $\epsilon= \pm 1$ to account for a canonical $(+)$ or phantom $(-)$ scalar field.

The corresponding field equations are obtained by varying the action (\ref{NEDaction}) with respect to $ A_\mu, \phi$, and $g^{\mu \nu}$, leading to:
\begin{align}
& \nabla_\mu\left[L_F F^{\mu \nu}\right]=\frac{1}{\sqrt{-g}} \partial_\mu\left[\sqrt{-g} L_F F^{\mu \nu}\right]=0, \label{fieldF} \\
& 2 \epsilon \nabla_\mu \nabla^\mu \phi=-\frac{d V(\phi)}{d \phi}, \\
& G_{\mu \nu}=R_{\mu \nu}-\frac{1}{2} g_{\mu \nu} R=\kappa^2\left(T_{\mu \nu}^\phi+T_{\mu \nu}^{E M}\right),
\end{align}
respectively, where $L_F=d L / d F$ whereas $T_{\mu \nu}^\phi$ and $T_{\mu \nu}^{E M}$ are the stress-energy tensor for the scalar field and electromagnetic field, respectively, which are given by:
\begin{align}
& T_{\mu \nu}^{E M}=g_{\mu \nu} L(F)-L_F F_\nu^\alpha F_{\mu \alpha}, \\
& T_{\mu \nu}^\phi=2 \epsilon\partial_\nu \phi \partial_\mu \phi-g_{\mu \nu}\left(\epsilon  \partial^\alpha \phi \partial_\alpha \phi-V(\phi)\right) .
\end{align}

In order to reconstruct the NED and scalar field densities we consider first the general case and next apply it to the modified SV space-time worked out in the previous section.
%Note. The definition in B here is inverted as compared to the equations.
We work in the line element (\ref{GeneralBB}). We will assume that our source has only electric charge, meaning that the only nonzero components of the electromagnetic tensor are $F^{tr}=-F^{rt}$. With that in mind, Eq.\eqref{fieldF} can be rewritten as
\begin{equation}
    \frac{1}{\sqrt{-g}} \partial_r\left[\sqrt{-g} L_F F^{rt}\right]=0
\end{equation}
which can be simply integrated as
\begin{equation}
\sqrt{\frac{A(r)}{B(r)}} C^2(r) L_F F^{tr}=q=\text{constant}\ .
\end{equation}
The invariant $F$ can be obtained by its definition as
\begin{equation}
    F=\frac{1}{4} F^{\mu \nu} F_{\mu \nu}=-\frac{1}{2}\frac{A(r)}{B(r)} (F^{tr})^2=-\frac{q^2}{2 C (r)^4L_F^2},
    \label{F}
\end{equation}
and the stress-energy tensors are given by:
\begin{align}
& T^{\phi^\mu}{ }_\nu=-\epsilon B(r) \phi^{\prime 2} \operatorname{diag}(1,-1,1,1)+\delta_\mu^\nu V(\phi), \\
& T^{E M^\mu}{ }_\nu=\operatorname{diag}\left(L+\frac{q^2}{L_F C(r)^4}, L+\frac{q^2}{L_F C(r)^4}, L, L\right)
\end{align}
Note that expressions for $\phi(r)$ and $L_F(r)$ can be found by a simple linear combination of the Einstein equations
\begin{align}
     & G^r {}_r - G^t {}_t=\kappa^2 (T^r {}_r - T^t {}_t)=2 \kappa^2 \epsilon \phi '(r)^2 B(r), \label{phi} \\
    & G^\theta {}_\theta - G^t {}_t=- \kappa^2 \frac{q^2}{C(r)^4 L_F}. \label{LF}
\end{align}
so after a given metric element is given, its Einstein tensor components can be computed and inserted in (\ref{phi}) and (\ref{LF}) in order to solve for the matter fields.

Similarly, the field equation of the scalar field $\phi$  allows us to obtain the potential $V(r)$ via the integration of the equation
\begin{align}
V'(r)=2 \epsilon \phi'(r)\sqrt{\frac{B(r)}{A(r)}} \frac{1}{C (r)^2} \partial_r \left ( \sqrt{A(r)B(r)} C (r)^2 \partial_r \phi(r) \right). \label{V}
\end{align}
Finally, the NED lagrangian is found in terms of the radial coordinate $r$ as
\begin{equation}
 G^r{ }_r+G^t{}_t=\kappa^2 T^r{ }_r+\kappa^2 T^t{}_t=2 \kappa^2\left(L+\frac{q^2}{L_F C(r)^4}+V\right) .   
\end{equation}
Should we be able to invert the relation in $F(r)$ and $\phi(r)$, we would arrive to the derivation of an explicit expression for the electromagnetic Lagrangian $L(F)$ and the scalar potential $V(r)$, though in practical terms this will only be possible in those cases in which the expressions above are simple enough.

\subsection{Modified SV case}

In the particular case of the modified SV space-time given by Eqs.(\ref{eq:SV1}) and (\ref{eq:SV2}), the scalar field equation \eqref{phi} provides the result
\begin{widetext}
\begin{equation}
    - \epsilon \kappa^2\phi'(r)^2=\frac{c^2}{(c^2+r^2)^2}\left[1+mr^2(c^2+r^2) \left( \frac{1}{\left(b^2+r^2\right) \left(\sqrt{b^2+r^2}-2 m\right)}-\frac{1}{\left(a^2+r^2\right) \left(\sqrt{a^2+r^2}-2 m\right)} \right )\right], \label{phipri}
\end{equation}
\end{widetext}
whose integration would be largely involved. This will be the case in general, in that these expressions cannot be typically integrated/inverted in order to get simple forms for the matter sources $\phi(r)$, $L(F)$ and the functions defining them, and one has to proceed on a case-by-case basis.

In order to deal with analytic solutions, let us focus on the case in which $a=b$, that is, when $A(r)=B(r)$. Under this restriction, the second term in Eq.\eqref{phipri} vanishes and the scalar field takes the following form
\begin{equation}
    \phi(r)=\frac{\tan^{-1} \left ( \frac{r}{c} \right )}{\kappa \sqrt{- \epsilon}}.
\end{equation}
Note that $\phi$ has to describe a phantom field (i.e. $\epsilon=-1$) since otherwise a solution for the scalar field cannot be found. On the other hand, it this case not every solution is mathematically consistent. To see why, we note that Eq.\eqref{V} can be integrated in this case to provide the expression:
\begin{widetext}
\begin{equation}
V(r)=\frac{2mc^2}{\kappa^2} \left ( \frac{3}{(c^2-a^2)^{\frac{5}{2}}} \tan^{-1} \left ( \sqrt{\frac{a^2+r^2}{c^2-a^2}} \right) +\frac{a^2+r^2+2(c^2+r^2)}{2(a^2-c^2)^2(c^2+r^2)\sqrt{a^2+r^2}} \right),
\end{equation}
\end{widetext}
which is only valid for $c>a$. 

Regarding the derivative of the NED Lagrangian, we can rewrite Eq.\eqref{LF} in the following way
\begin{equation}
    L_F(r)=\frac{\kappa^2 q^2\left(a^2+r^2\right)^{5 / 2}}{m\left(c^2+r^2\right)\left(2 c^2 r^2-2 a^4-a^2\left(c^2+5 r^2\right)\right)}.
\end{equation}
which in combination with the results above allows to find the NED Lagrangian in this case as
\begin{widetext}
\begin{align}
    L(r)=\frac{m}{\kappa ^2} \Bigg [ \frac{2a^4-2c^2r^2+a^2(c^2+5r^2)}{(a^2+r^2)^{5/2}}-\frac{2a^2}{(a^2+r^2)^{3/2} (c^2+r^2)} + \nonumber \\ 2c^2 \left ( \frac{3}{(c^2-a^2)^{\frac{5}{2}}} \tan^{-1} \left ( \sqrt{\frac{a^2+r^2}{c^2-a^2}} \right) +\frac{a^2+r^2+2(c^2+r^2)}{2(a^2-c^2)^2(c^2+r^2)\sqrt{a^2+r^2}} \right) \Bigg].
\end{align}
\end{widetext}
The next step is to invert the expressions for $\phi(r)$ and $F(r)$ in order to obtain those for $V(\phi)$ and $L(F)$. The former can be done easily enough to provide the expression
\begin{widetext}
\begin{equation}
V(\phi)=\frac{m}{\kappa^2}\left(\frac{6 c^2}{\left(c^2-a^2\right)^{5 / 2}} \tan^{-1}\left(\frac{\sqrt{a^2+c^2 \tan(\sqrt{-\epsilon} \kappa \phi)^2}}{\sqrt{c^2-a^2}}\right)+\frac{a^2+5 c^2+(a-c)(a+c) \cos(2 \sqrt{-\epsilon} \kappa \phi)}{\left(a^2-c^2\right)^2 \sqrt{a^2+c^2 \tan(\sqrt{-\epsilon} \kappa \phi)^2}}\right).
\end{equation}
\end{widetext}
while for the former it turns out that $F(r)$ is not invertible. We can nonetheless circumvent this problem by defining an auxiliary field $P^{\mu \nu}=L_F F^{\mu\nu}$, whose general expression reads
\begin{equation}
    P=\frac{1}{4}P^{\mu \nu} P_{\mu \nu}=-\frac{q^2}{2 C (r)^4},
\end{equation}
in terms of which the Lagrangian density in the present case becomes
\begin{widetext}
\begin{align}
&L(P)= \frac{2m}{\kappa^2} \Bigg[-\frac{4 a^2}{\sqrt{P}q\left(2 a^2-2 c^2+\sqrt{2} \sqrt{P}q\right)^{3 / 2}}+\frac{4 \sqrt{2}\left(a^2-c^2\right)^2+10 a^2 \sqrt{P}q-4 c^2 \sqrt{P}q}{\left(2 a^2-2 c^2+\sqrt{2} \sqrt{P}q\right)^{5 / 2}} \nonumber\\
& +c^2\left(\frac{-2 a^2+2 c^2-3 \sqrt{2} \sqrt{P}q}{\left(a^2-c^2\right)^2 \sqrt{P}q \sqrt{2 a^2-2 c^2+\sqrt{2} \sqrt{P}q}}-\frac{3}{\left(c^2-a^2\right)^{5 / 2}} \tan^{-1}\left(\frac{\sqrt{a^2-c^2+\frac{\sqrt{P}q}{\sqrt{2}}}}{\sqrt{c^2-a^2}}\right)\right)\Bigg].
\end{align}
\end{widetext}

%%%%%%%%%%%%%%%%%%%
\section{Energy Conditions}
\label{energy}
%%%%%%%%%%%%%%%%%%%%%

Let us now analyze the energy conditions for the modified SV space-time. We consider only the null energy condition (NEC) since if it is violated then any other energy conditions will be violated too. We consider an anisotropic fluid with density $\rho$ and radial and tangential pressures in order to accommodate the features of the stress-energy tensor of the NED. Outside any horizon that may be present, we have $T^t{}_t=-\rho$ and $T^r{}_r=p_\parallel$, while inside the horizon (if there is so), we have $T^t{}_t=p_\parallel$ and $T^r{}_r=-\rho$. Therefore, the null energy condition reads:
\begin{widetext}
\begin{equation}
    E_\text{null}=\rho+p_\parallel =\begin{cases}
    -T^t{}_t+T^r{}_r=-\frac{2A(r)}{\kappa^2C(r)}\sqrt{\frac{B(r)}{A(r)}}\partial_r \left( \sqrt{\frac{B(r)}{A(r)}}C'(r) \right) \quad \quad &\text{outside the event horizon} \\ \\
    T^t{}_t-T^r{}_r=\frac{2A(r)}{\kappa^2C(r)}\sqrt{\frac{B(r)}{A(r)}}\partial_r \left( \sqrt{\frac{B(r)}{A(r)}}C'(r) \right) \quad \quad&\text{inside the event horizon}
    \end{cases}
    \label{NEC}
\end{equation}
\end{widetext}
Note that for the case $A(r)=B(r)$, Eq.\eqref{NEC} reduces to the results of Ref. \cite{Alencar:2024yvh}, something that also happens when we set $a=b$, since we recover the original SV metric. In the modified SV metric considered in this work, outside the horizon the null energy condition would be satisfied as far as the following condition holds:
\begin{equation}
    -\partial_r \left( \sqrt{\frac{B(r)}{A(r)}}C'(r) \right)\geq0 \rightarrow  \frac{C''(r)}{C'(r)}\leq \frac{B(r)}{2A(r)} \partial_r \left( \frac{A(r)}{B(r)} \right)\ ,
    \label{cond}
\end{equation}
where we have used the fact that away from the horizon $A(r)\geq 0$ and $B(r)\geq 0$. Note that the left-hand side of \eqref{cond} is always positive, therefore, the right-hand side has to be, at least, positive in order to satisfy the null energy condition. Hence, the NEC is violated everywhere outside the horizon when $a>b$. Finally, in the case in which $b>a$, the limit $r\rightarrow \infty$ of the left-hand side approaches $0$ as $r\sim 1/r^3$, whereas the right-hand side approaches $0$ as $r\sim 1/r^4$, such that there will be always a region where the inequality is not satisfied, implying that the null energy condition will be always violated, as expected.

%The study of the null energy condition can be performed directly form Eq. (\ref{phi}). Since $B$ always changes sign when the horizon is crossed, Eq. (\ref{phi}) implies that the null energy condition is always satisfied.

%%%%%%%%%%%%%%%%%%%%%%%
\section{Conclusion}
%%%%%%%%%%%%%%%%%%%%%%%

In this paper a general procedure for constructing spherically symmetric black bounces is provided. We have shown that by departing from a general spherically symmetric metric with three independent components, the metric can be reduced just to two independent functions by following the appropriate coordinate transformation, a usual procedure for this type of metrics. Nevertheless, depending on the existence of roots in the $tt$ and/or the $rr $ components of the line element, the metric might not behave well despite the initial conditions imposed on such components even after regarding their smoothness over all the space and the asymptotically flatness character of such a metric. We classified the different combination of roots into five different cases, and discussed their respective interpretations and the change of coordinates needed to remove any (if possible) artificial coordinate singularity.

We next applied such a procedure to a generalization of the Simpson-Visser space-time, the latter being the original black bounce solution, via an additional parameter so that the final metric depends on two arbitrary parameters, besides the asymptotic mass $M$.  We found that depending on the relative value of the free parameters, the solution may interpolate between a regular black hole and a traversable wormhole, as in the original Simpson-Visser space-time, or it may lead to an ill-defined metric where a conic singularity arises at the origin. 

In order to reconstruct the corresponding matter Lagrangian, a combination of non-linear electrodynamics with a scalar field is required, as usual for black bounce solutions, since the radial and tangential pressures required to support the bounce differ (and thus it cannot be reproduced with NEDs alone). Nevertheless, a general procedure is provided such that any spherically symmetric black bounce can be reconstructed by just introducing the appropriate components of the metric. In addition, by analyzing the energy conditions, we found that the null energy conditions is always violated at least in some region of the radial coordinate, in agreement with the hypothesis of the singularity theorems for any regular solution within GR.

Our results show that implementations of the black bounce idea must take special care on the zeros of the metric components and the suitable change of coordinates to bring the corresponding line element into a well-behaved form and provide a suitable interpretation in terms of regular black holes, traversable wormholes, or solutions with conical singularities. Furthermore, for every such solution the corresponding NED plus scalar field models supporting it can be re-constructed, while the energy conditions will always be violated in some region. These results may serve as a starting point to build black bounces framed within extensions of General Relativity, where the energy conditions are not necessarily violated in order to implement the bounce and its regularity. Work on this direction is currently underway.

\section*{Acknowledgements}

The authors would like to thank Conselho Nacional de Desenvolvimento Científico e Tecnológico (CNPq), Fundação Cearense de Apoio ao Desenvolvimento Científico e Tecnológico
(FUNCAP) and Coordenação de Aperfeiçoamento de Pessoal de Nível Superior - Brasil (CAPES) for finantial support. Financial support of the Department of Education, Junta de Castilla y León, and FEDER Funds is also gratefully acknowledged (Reference: CLU-2023-1-05). This work is supported by the Spanish grants Ref.~PID2020-117301GA-I00 (DS-CG and ADC) and PID2022-138607NBI00 (DRG) funded by MCIN/AEI/10.13039/501100011033 (``ERDF A way of making Europe" and ``PGC Generaci\'on de Conocimiento").

\end{document}